# Interaction of (-)-epigallocatechin gallate with silver nanoparticles

Goutam Kumar Chandra[1], Debi Ranjan Tripathy[2], Swagata Dasgupta[2] and Anushree Roy[1*]

[1]Department of Physics, Indian Institute of Technology Kharagpur, 721302, India

[2]Department of Chemistry, Indian Institute of Technology Kharagpur, 721302, India

## Abstract

Interactions between silver nanoparticles and (-)-epigallocatechin gallate (EGCG) have been investigated. Prior to the addition of EGCG molecules the silver particles are stabilized by borate ions. Studies on the surface plasmon resonance band of silver particles suggest that the EGCG molecules remove the borate ions from the surface of the metal particles due to the chelating property of the ions. The complex formation by EGCG and borate ions has been confirmed by NMR studies and pH titration. A possible scheme of interaction between the two has been proposed.

Electronic mail: anushree@phy.iitkgp.ernet.in

## 1. Introduction

Polyphenols are natural substances, present in beverages obtained from plants, fruits and vegetables, such as red wine and tea. The dominant and most important catechin in green tea is (-)-Epigallocatechin gallate (EGCG), a potent antioxidant, which has been reported to show anti-cancer activity and is presently under investigation in clinical studies [1-3]. Silver ions ($Ag^+$) and silver-based compounds are also known to exhibit strong antimicrobial effects [4, 5]. The active ingredients in silver-based drugs are "oligodynamic" silver ions [6]. In medical literature, an interesting case-history reveals the oncolytic role of $Ag^+$ [7]. This case history also reports the role of EGCG in anti-cancer dietary therapy. However, $Ag^+$ can have only limited usefulness as an antimicrobial agent due to the interfering effects of their salts [8]. Such adverse effects of $Ag^+$ can be avoided by using the corresponding metal nanoparticles. Use of silver (Ag) nanoparticles is an efficient and reliable tool for improving the biocompatibility of Ag in different biological processes. It is common knowledge that the biological activities of several polyphenols are affected by transition metal ions [9-11]. With increased use of Ag nanoparticles as drug carriers it would be interesting to investigate its interaction with EGCG.

## 2. Experimental

Silver colloids are prepared by using sodium borohydride ($NaBH_4$) as a reducing agent. Silver nitrate ($AgNO_3$), sodium borohydride ($NaBH_4$) of analytical reagent grade (SRL, India), were used to prepare the silver sol. A colloidal silver solution was prepared in deionized water following the method described by Creighton et al [12]. In this chemical route, $AgNO_3$ is reduced by an excess amount of $NaBH_4$. 2.2 mM $AgNO_3$ was added dropwise to 1 mM $NaBH_4$ at 4°C. Vigorous stirring for 20 minutes was necessary to stabilize the colloidal solution. Later, it was left at room temperature for approximately 1 hour till the solution became transparent yellow. EGCG was obtained from Sigma-Aldrich (USA).

UV-Visible (UV-Vis) spectra were recorded using a spectrophotometer, Model Lambda-45 (Make Perkin-Elmer). For optical absorption measurements the samples were kept in a microcuvette with an optical pathlength of 1 cm. $^1$H NMR spectra were recorded on a Bruker 400 MHz spectrometer at 22.5 °C.

## 3. Results and Discussion

### 3.1 Formation of borate esters of EGCG

The pH of the Ag sol is measured to be 8.43, which remains constant during the time course of the experiment. Just after addition of 0.5 mM of EGCG, the pH of colloidal solution decreases to 7.72. As time progresses the pH of the solution gradually decreases further. After 8 hrs the pH of the solution reaches a value ~ 7.30 and remains constant with time as shown in Fig. 1. The acid dissociation constants of EGCG molecules (inset of Fig. 1) [13] indicate that in a basic medium deprotonation of the aromatic OH groups of the molecule occurs. The decrease in pH of the colloidal solution after addition of EGCG, thus, indicates the release of $H^+$ due to such an ad-molecule in the sol. We believe that the pH of the solution decreases further due the formation of borate esters of EGCG [14].

The complexation of EGCG with borate ion has been demonstrated by NMR spectroscopic measurements. The $^1$H NMR spectrum of the gallate and pyrogallol moiety of EGCG in $D_2O$ in water suppression mode are (δ6.883(s)2H) and (δ6.474(s)2H) (shown in Fig. 2(a)). In presence of NaOH the $^1$H NMR spectrum of gallate and pyrogallol moiety of EGCG in $D_2O$ in water suppression mode are (δ6.869(s)2H) and (δ6.469(s)2H) (shown in Fig. 2(b)). However, in presence of $NaBH_4$ the $^1$H NMR spectrum of gallate and pyrogallol moiety of EGCG in $D_2O$ in water suppression mode are (δ6.878(s) and δ6.748(s) 2H) and (δ6.467(S) and δ6.345(s) 2H) (shown in Fig. 2(c)). Due to the formation of a chelate involving the ortho phenolic group with borate, there is a split in the NMR spectrum along with broadening [15-16]. We have seen that two protons show more shielding which is only possible if boron is tetra coordinated and with a negative charge. Since the two aromatic protons of ring A are deuterated they do not appear here. The possible path of the reaction of EGCG and borate ions is proposed in Scheme 1.

**3.2 Stability of metal colloids with EGCG in solution**

Furthermore, studies on the surface plasmon resonance (SPR) band of the Ag particle provide direct evidence of the interaction between Ag colloids and EGCG. The SPR band of Ag colloids (shown in Fig. 3) appears at 397 nm. The measurements were carried out immediately after direct addition of EGCG molecules (0.5 mM in the sol) in the metal sol and subsequently at half hour intervals over a duration of 3 hrs. The broken lines in Fig. 3 demonstrate the change in spectral profile of the plasmon band of Ag colloids at different time intervals on addition of EGCG (for clarity, we have shown only a few characteristic spectra taken just after mixing (0 hr), after 1 hr, 2 hrs and 3 hrs). We observe (i) an increase in intensity of the absorption maximum and (ii) a red shift (by 8 nm) of the absorption maximum of plasmon band on addition of EGCG in the beginning, which remains constant at a later time.

The nanosized particles possess a very large surface-to-volume ratio, and consequently their properties are mostly governed by the surface states. It is well known that the SPR band of Ag colloids in solution is strongly influenced by any chemical modification of the surface, depending on whether the ad-molecule is nucleophilic and therefore donates electron density into the particle surface (blue-shift of the SPR band), or is electrophilic and withdraws electron density from the particle surface (red-shift of the SPR band). In the present case, the Ag particles in the sol are stabilized by the anions (borate ions). Thus the red-shift of the SPR band of the Ag colloids (shown in Fig. 3) by addition of EGCG indicates a decrease of electron density (anions) from the colloidal surface that indicates a desorption or removal of anions (borate ions) from the surface of the metal particles.

As the Ag colloids in our experiments are suspended in an aqueous solution and excess $NaBH_4$ was used in the chemical synthesis, the following experiments were performed to obtain a clearer view on the metal-molecule interaction. The absorption band of an aqueous solution of EGCG appears at 275 nm (solid line in Fig. 4(a)) whereas, a double peak absorption band of 0.5 mM of EGCG in 1 mM $NaBH_4$ (same concentration as added during synthesis of Ag sol) solution appears at 279 nm and 316 nm (shown by dotted line in Fig. 4(a)). This feature of the absorption band of EGCG in $NaBH_4$ solution, most likely arises from the deprotonated EGCG molecule and the corresponding borate esters of EGCG. The absorption spectra obtained on successive addition of 0.5 mM aqueous solution of EGCG and $NaBH_4$ solution of EGCG to Ag colloids is given in Fig. 4(b) and Fig. 4(c), respectively. On addition of an aqueous solution of EGCG to Ag colloids, the characteristic absorption band at 275 nm of EGCG remains unchanged and another weak shoulder originating from the EGCG-borate complex appears at a higher wavelength (indicated by an arrow in Fig. 4(b)). Initially we observe an increase in intensity and a red shift of the plasmon band of the metal colloid by ~ 8 nm. Subsequently, there is a further red shift of the plasmon band (as shown in Fig. 4(b)) for the higher concentration of the added molecule. However, on addition of $NaBH_4$ solution of EGCG in Ag sol, along with the appearance of the characteristic double peak structure of deprotonated EGCG and borate ester in the spectra, there is an increase in intensity and red-shift by ~ 4 nm of the absorption maximum of plasmon band of silver colloids (Fig. 4(c)). The relative shift of the plasmon band in above two cases has been compared in Fig. 5. The smaller shift in plasmon band in Fig. 4(c) than what was observed in Fig. 4(b) (shown by filled and open circles, respectively, in Fig. 5), can be explained by the fact that in case of the aqueous solution of EGCG in Ag colloids, the ad-molecules are free to strip off the borate ions present at the surface of the Ag colloids and can form a complex. However, for EGCG dissolved in $NaBH_4$ solution the molecules are deprotonated and form a complex with borate ions (prior to addition to the colloidal solution of Ag). Thus, the probability of the EGCG molecules to interact with the borate ions present on the surface of the Ag colloids is lower in comparison to the previous case where the interaction of EGCG occurs directly with the sol. The above results, once again, indicate that EGCG molecules interact with the borate ions on the surface of Ag particles.

### 3.3 Change in surface charge of Ag colloids in presence of EGCG

Next we estimate the change in surface charge and surface potential of Ag colloidal particles due to the interaction with the EGCG molecules. The shift in the wavelength of absorption maximum of the plasmon band before ($\lambda_0$) and after ($\lambda$) desorption of the anions from the particle surface is given by the equation [17],

$$\lambda = \lambda_0 \left(1 + \frac{[\text{anion}]}{[\text{Ag}]}\right)^{1/2} \qquad \text{........................ (1)}$$

where [Ag] is the net silver concentration in the colloid (distributed in all particles of the colloidal solution) and [anion] is the concentration of anions removed from the particle surface. The spectral shift ($\lambda$-$\lambda_0$) in Fig. 3 is 61 meV. Hence, we estimate the fractional change in charge density upon desorption of anions from the surface of the Ag particles to be $\Delta q = \frac{[\text{anion}]}{[\text{Ag}]}$ = 0.04, which is equivalent to ~ $0.5\times10^{-4}$ M concentration of borate ions (anion) in solution for [Ag]=$1.2\times10^{-3}$ M (used during sample preparation).

To estimate the change in surface potential of the Ag particles due to desorption of anions from the surface by EGCG, we have studied the change in maximum wavelength of the SPR band of Ag colloids with different concentrations of EGCG in the solution (shown in Fig. 6). From the shift in the plasmon band and using the expression in Eq. 1, we estimate the fractional change in charge density of metal particles with increase in concentration of EGCG in solution. The change in surface potential arising from the change in charge $\Delta q$ on the surface of particle of radius R as estimated using the expression $\Delta V = \Delta q/4\pi\varepsilon R$ [18] is shown in inset of Fig. 6. It is clear that as we increase the concentration of EGCG there is an initial increase in $\Delta V$. At around 0.2 to 0.3 mM concentration of EGCG, $\Delta V$ reaches a maximum value and then decreases with further increase in the concentration of the ad-molecule. A possible explanation to the observed change in surface potential of the Ag colloid on addition of EGCG is given. With increase in concentration, as more EGCG molecules are available in the solution, higher amounts of borate ions are removed from the surface of Ag colloids and thus, the surface potential of the colloids increases as shown in Fig. 6. However, with further increase in concentration of EGCG, the excess ad-molecules (after removing the negative ions layers from the colloidal surface) or the complex start to interact directly with the surface of the colloids (by neutralizing the particles) which decreases the surface potential of the particle.

Spectroscopic measurements thus indicate that the added EGCG molecules are capable of stripping off borate ions from the surface of the Ag colloidal particles in the sol with the formation of borate esters. This study has far reaching implications in terms of the usage of metal nanoparticles as drug carriers. Changes in chemical composition and surface charge for such drug-nanoparticle combinations require further investigation to be able to use them effectively and beneficially.

## Acknowledgments

A. Roy and S. Dasgupta thank DRDO, India for financial assistance. D. R. Tripathy is thankful to CSIR, India for a Junior Research Fellowship.

## Figure Captions

**Figure 1.** Change in pH of Ag colloids on addition of EGCG in the sol. Inset presnts the structure of the EGCG molecule and the pKa value of two protons are shown.

**Figure2.** (a) 1H NMR spectrum of EGCG in $D_2O$ in water suppression mode. (b) 1H NMR spectrum of EGCG and NaOH in $D_2O$ in water suppression mode, (c) 1H NMR spectrum of EGCG and $NaBH_4$ in $D_2O$ in water suppression mode.

**Figure 3.** Time dependent study of the plasmon band of Ag colloids on addition of EGCG in the sol.

**Figure 4.** (a) The solid and dash lines represent the absorption spectra of EGCG in aqueous solution and EGCG in $NaBH_4$ solution respectively. (b) Titration of Ag colloids with 0.5 mM EGCG in aqueous solution. (c) Titration of Ag colloids with 0.5 mM EGCG in $NaBH_4$ solution. The red arrows indicate the order of the spectra taken with increase in volume ratio of EGCG.

**Figure 5.** Comparison between the relative shift of the plasmon band in case of aqueous solution of EGCG (filled circles) and $NaBH_4$ solution of EGCG (open circles) in Ag colloids.

**Figure 6.** Concentration dependent study of the plasmon band of Ag colloids on addition of different concentration of EGCG in the Ag sol. Inset of the figure shows the change in surface potential of the Ag colloids ($\Delta V$) with different concentration (C-EGCG) of EGCG.

**Scheme 1.** Complex formation of EGCG with $NaBH_4$.

## Figures

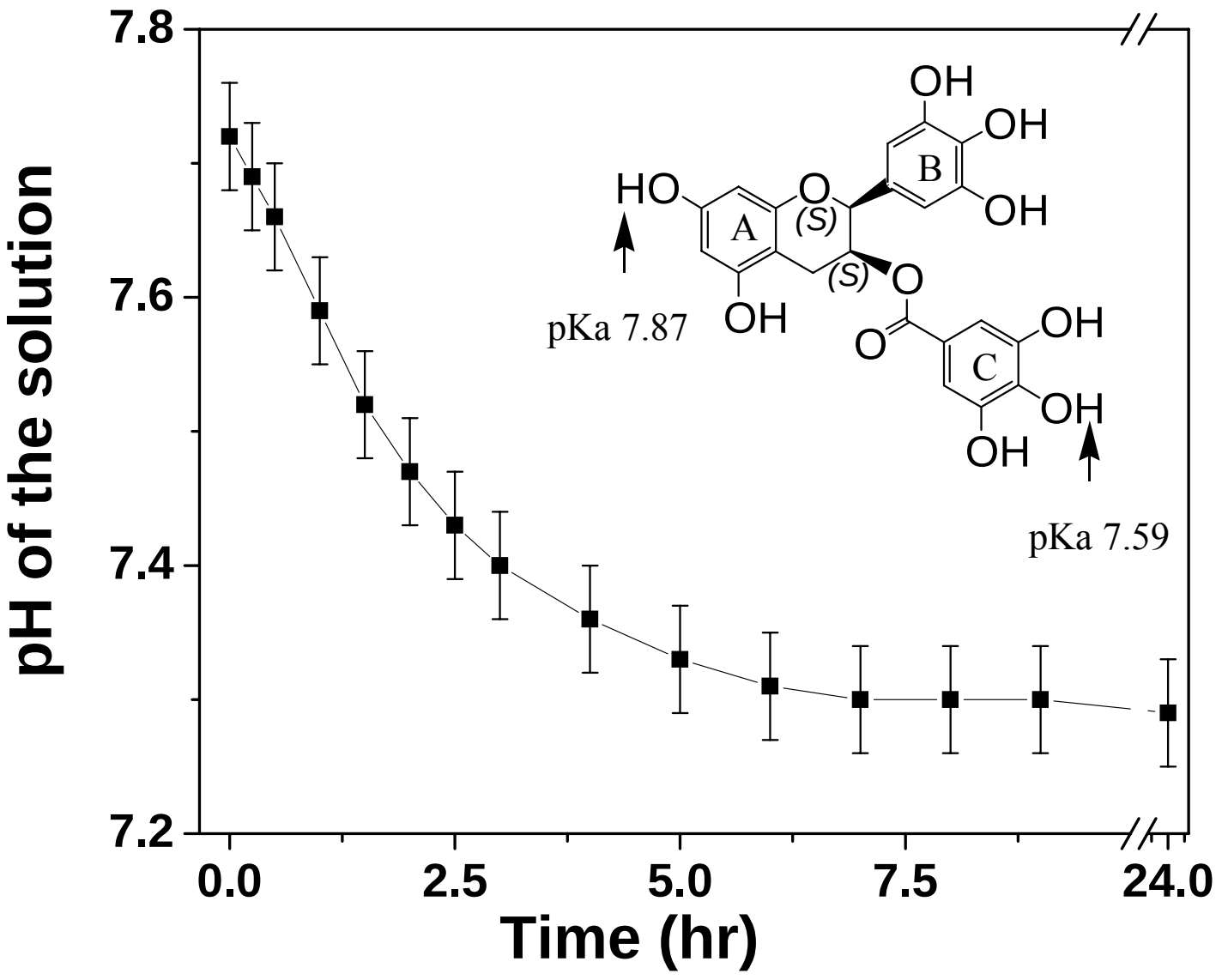

Figure 1

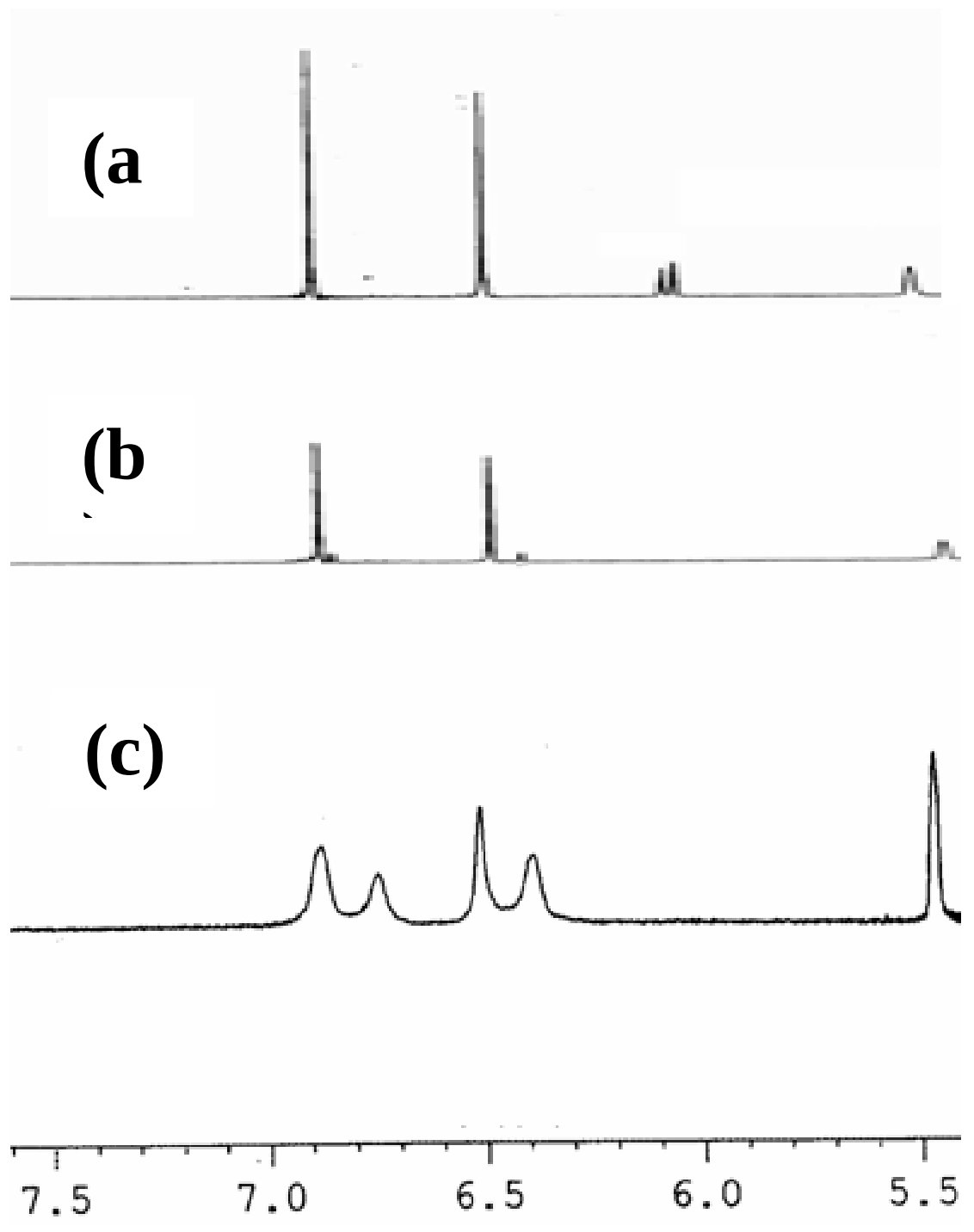

Figure 2

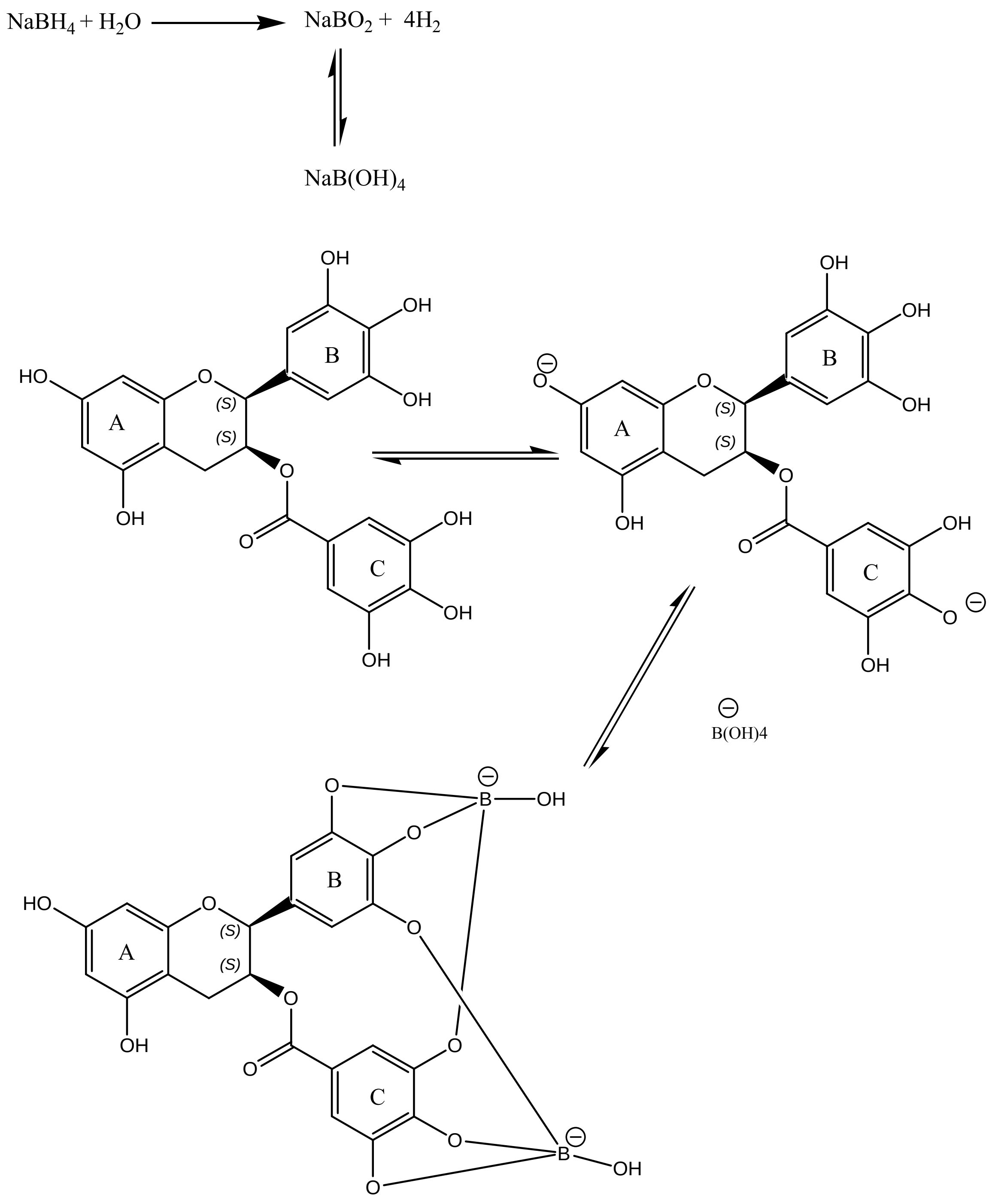

**Scheme 1**

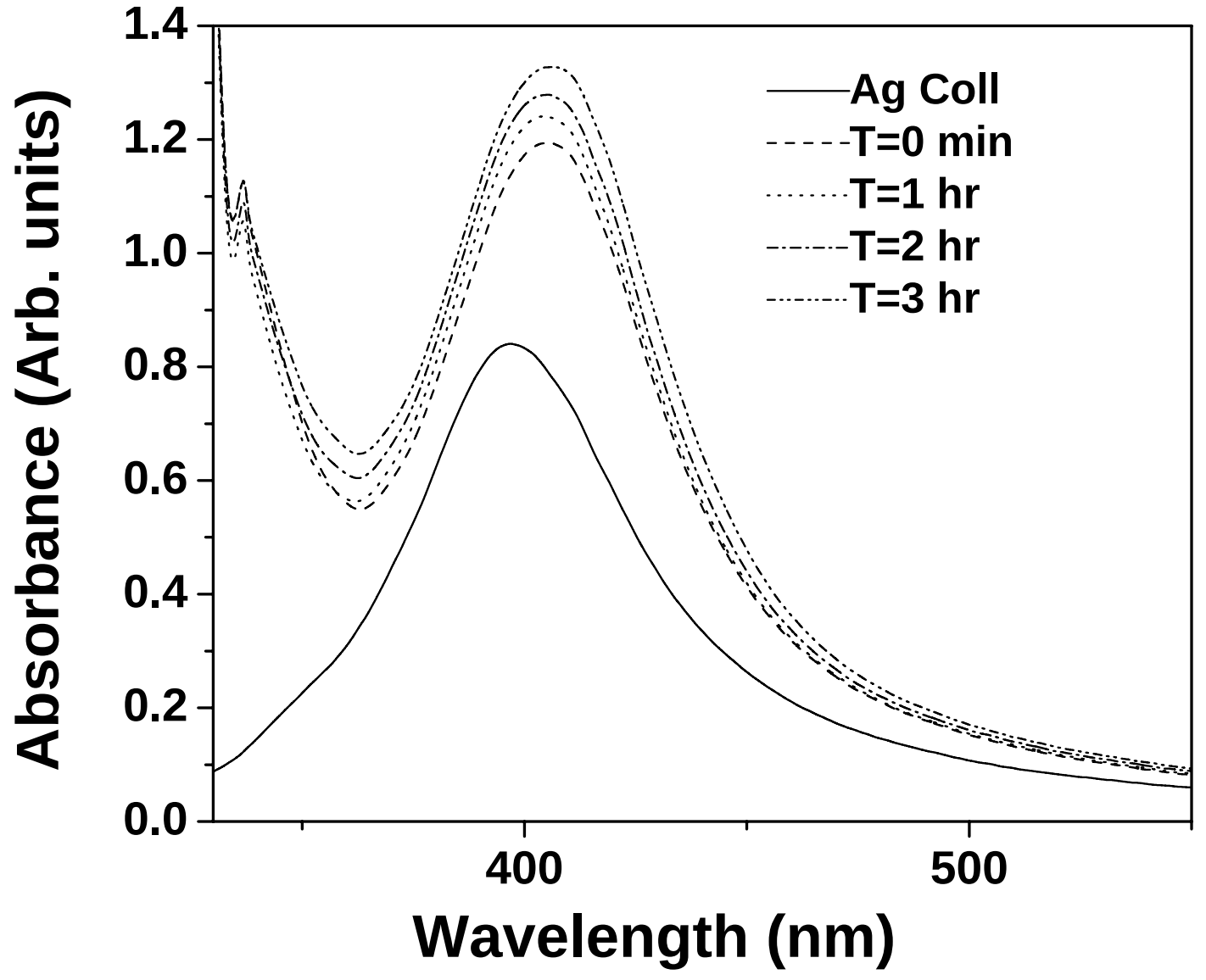

**Figure 3**

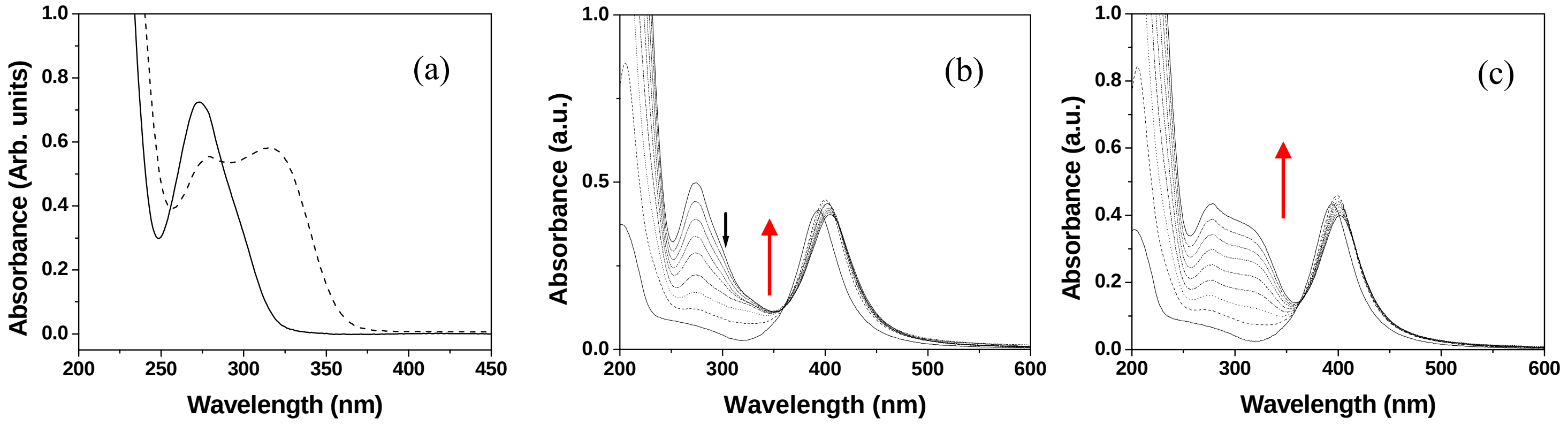

**Figure 4**

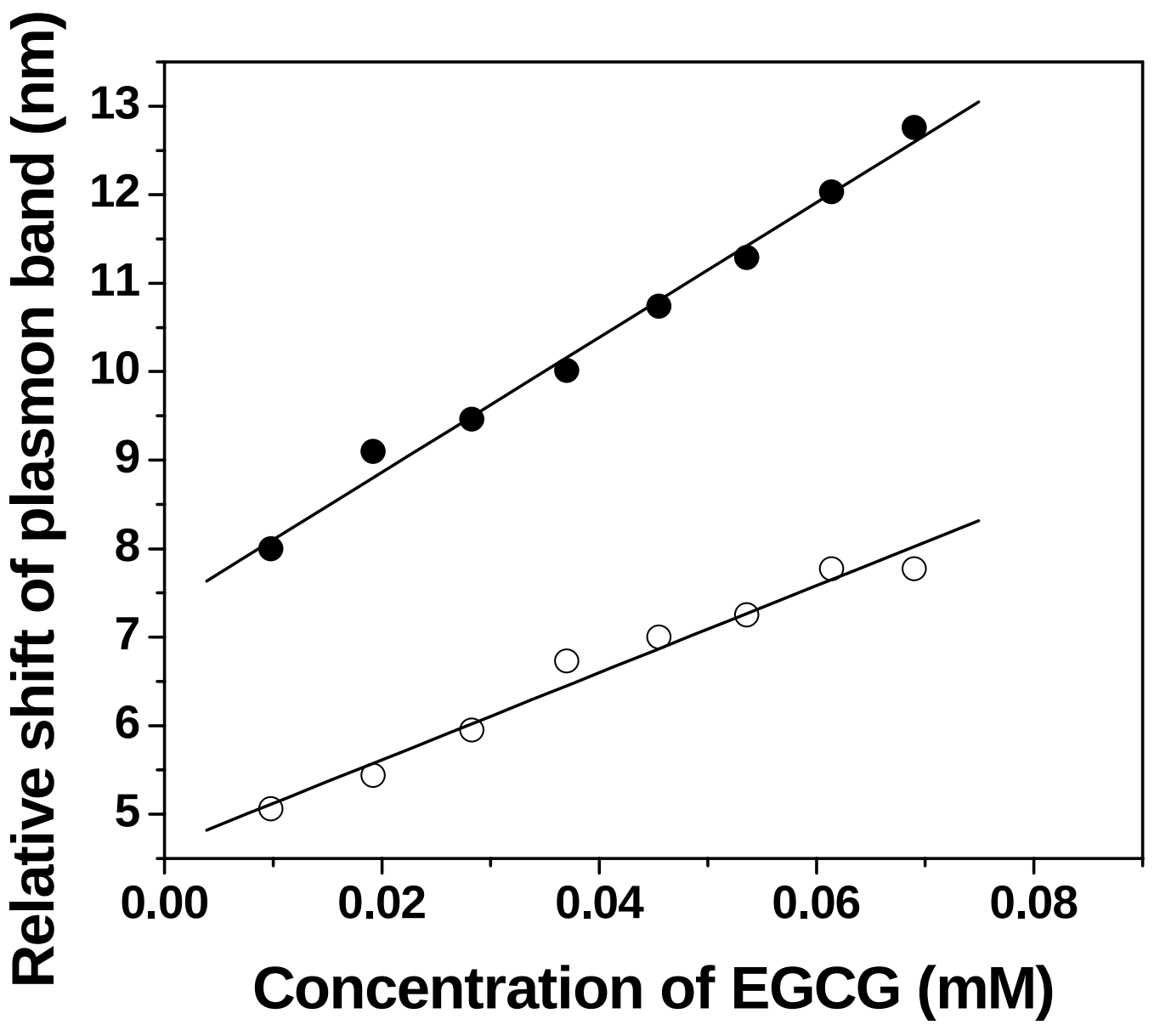

**Figure 5**

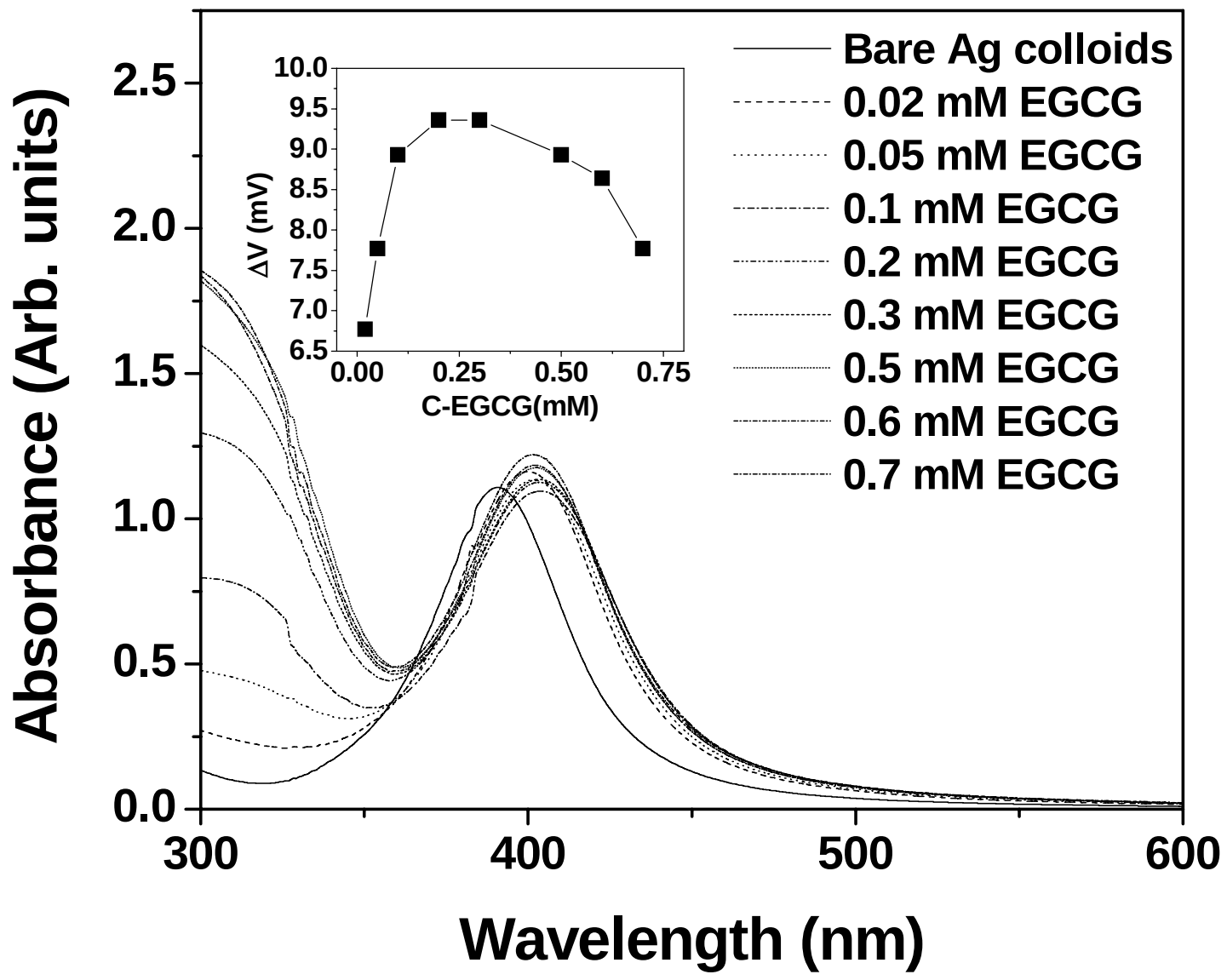

**Figure 6**